\documentstyle[11pt,epsfig,epsf]{article}

\newlength{\dinwidth}
\newlength{\dinmargin}
\def\lapproxeq{\lower .7ex\hbox{$\;\stackrel{\textstyle <}{\sim}\;$}}
\def\gapproxeq{\lower .7ex\hbox{$\;\stackrel{\textstyle >}{\sim}\;$}}

\def\be{\begin{equation}}
\def\ee{\end{equation}}
\def\bea{\begin{eqnarray}}
\def\eea{\end{eqnarray}}

\def\lesim{ \;\raisebox{-.7ex}{$\stackrel{\textstyle <}{\sim}$}\; }

\def\ra{ \rightarrow }

\begin{document}
\titlepage
\begin{flushright}
KEK--TH--844\\
IPPP/02/52 \\
DCPT/02/104\\
CERN--TH/2002-233\\
20 September 2002 \\
\end{flushright}

\vspace*{4cm}

\begin{center}
{\Large \bf The SM prediction of $g-2$ of the muon}

\vspace*{1cm} {\sc K. Hagiwara}$^a$, {\sc A.D. Martin}$^b$, {\sc
Daisuke Nomura}$^b$,
and {\sc T. Teubner}$^c$ \\

\vspace*{0.5cm}
$^a$ Theory Group, KEK, Tsukuba, Ibaraki 305-0801, Japan \\
$^b$ Department of Physics and Institute for
Particle Physics Phenomenology,\\
University of Durham, Durham DH1 3LE, U.K. \\
$^c$ Theory Division, CERN, CH-1211 Geneva 23, Switzerland
\end{center}

\vspace*{1cm}

\begin{abstract}
We calculate $(g-2)/2$ of the muon, by improving the determination of
the hadronic vacuum polarisation contribution, $a_\mu^{\rm
had,LO}$, and its uncertainties. The different $e^+e^-$ data sets
for each exclusive (and the inclusive) channel are combined in
order to obtain the optimum estimate of the cross sections and
their uncertainties. QCD sum rules are evaluated in order to
resolve an apparent discrepancy between the inclusive data and the
sum of the exclusive channels. We conclude $a_\mu^{\rm
had,LO}=(683.1\pm5.9_{\rm exp} \pm 2.0_{\rm rad})\times10^{-10}$
which, when combined with the other contributions to $(g-2)/2$, is
about $3\sigma$ below the present world average measurement.
\end{abstract}

\newpage
\section{Introduction}

The muon anomalous magnetic moment, $a_\mu \equiv (g_\mu - 2)/2$,
is one of the most precisely measured quantities in contemporary
particle physics.  The world average of the existing measurements
is
\begin{eqnarray}
 a_\mu^{\rm exp} = 11 659 203(8) \times 10^{-10},
 \label{eq:BNL2001}
\end{eqnarray}
which is dominated by the recent value obtained by the E821
collaboration at BNL\cite{BNL2002}.
It is so precisely measured that it is very useful in probing and
constraining New Physics beyond the Standard Model (SM). It is
therefore important to evaluate the SM prediction of $a_\mu$ as
accurately as possible.

The SM contribution to $a_\mu$ may be written as the sum
of three terms,
\begin{eqnarray}
 a_\mu^{\rm SM} = a_\mu^{\rm QED} + a_\mu^{\rm EW} + a_\mu^{\rm had} .
\label{eq:amusm}
\end{eqnarray}
The QED contribution, $a_\mu^{\rm QED}$, has been calculated up to
and including estimates of the 5-loop contribution, see reviews
~\cite{HuK,CM},
\begin{eqnarray}
 a_\mu^{\rm QED} = 116~584~705.6 (2.9) \times 10^{-11} .
 \label{eq:QED}
\end{eqnarray}
In comparison with the experimental error in eq.~(\ref{eq:BNL2001}), and with
the hadronic contribution error discussed later, the uncertainty
in $a_\mu^{\rm QED}$ is much less important than other sources of
uncertainty.
The electroweak contribution $a_\mu^{\rm EW}$ is
calculated through second order to be~\cite{CKM, PPdR, KPPdR}
\begin{eqnarray}
 a_\mu^{\rm EW} = 152 (1) \times 10^{-11} .
 \label{eq:EW}
\end{eqnarray}
Here again the error is negligibly small.

Less accurately known is the hadronic contribution $a_\mu^{\rm
had}$. It can be divided into three pieces,
\begin{eqnarray}
 a_\mu^{\rm had} = a_\mu^{\rm had,LO}
                 + a_\mu^{\rm had,NLO}
                 + a_\mu^{{\rm had,l}\raisebox{0.45ex}{\rule{0.6ex}{0.08ex}}{\rm b}\raisebox{0.45ex}{\rule{0.6ex}{0.08ex}}{\rm l}}.
\end{eqnarray}
The lowest-order (vacuum polarisation) hadronic contribution,
$a_\mu^{\rm had,LO}$, has been calculated by a number of groups.
The value
\begin{eqnarray}
 a_\mu^{\rm had,LO} = 6~924 (62) \times 10^{-11}
 \label{eq:hadLO_byDH98b}
\end{eqnarray}
taken from Ref.~\cite{DH98b} has been frequently used in making
comparisons with the data. The next-to-leading order hadronic
contribution, $a_\mu^{\rm had,NLO}$, is evaluated to
be~\cite{Krause,ADH98}
\begin{eqnarray}
  a_\mu^{\rm had,NLO} = -100 (6) \times 10^{-11}.
  \label{eq:hadNLO_byKrause}
\end{eqnarray}
%
%
The hadronic light-by-light scattering contribution $a_\mu^{{\rm
had,l}\raisebox{0.45ex}{\rule{0.6ex}{0.08ex}}{\rm
b}\raisebox{0.45ex}{\rule{0.6ex}{0.08ex}}{\rm l}}$ has been
recently reevaluated~\cite{KN}--\cite{RMW}, and it is found to be
\begin{eqnarray}
  a_\mu^{{\rm had,l}\raisebox{0.45ex}{\rule{0.6ex}{0.08ex}}{\rm b}\raisebox{0.45ex}{\rule{0.6ex}{0.08ex}}{\rm l}} = 80 (40) \times 10^{-11} ,
  \label{eq:hadlbyl}
\end{eqnarray}
where we quote the estimate of the full hadronic light-by-light
contributions given in Ref.~\cite{Nyffeler}. From
eqs.~(\ref{eq:hadLO_byDH98b}), (\ref{eq:hadNLO_byKrause}) and
(\ref{eq:hadlbyl}), we can see that $a_\mu^{\rm had,LO}$ has the
largest uncertainty, although the uncertainty in the
light-by-light contribution $a_\mu^{{\rm
had,l}\raisebox{0.45ex}{\rule{0.6ex}{0.08ex}}{\rm
b}\raisebox{0.45ex}{\rule{0.6ex}{0.08ex}}{\rm l}}$ is also large.

In this letter we update the evaluation of $a_\mu^{\rm had,LO}$,
which is given by the dispersion relation
\begin{eqnarray}
 a_\mu^{\rm had,LO}
=
 \frac{1}{4\pi^3} \int_{s_{\rm th}}^\infty ds\ \sigma_{\rm had}(s)\left(\frac{m_\mu^2}{3s}\ K(s)\right),
 \label{eq:dispersion_rel}
\end{eqnarray}
where $\sigma_{\rm had}(s)$ is the total cross section for
$e^+e^-\to {\rm hadrons}\,(+\gamma)$ at centre-of-mass energy
$\sqrt{s}$. The kernel function $K(s)$ is given by
\begin{eqnarray}
 K(s>4m_\mu^2) = \frac{3s}{m_\mu^2}\left\{\frac{x^2}{2}(2-x^2) + \frac{(1+x^2)(1+x)^2}{x^2}
        \left( \ln ( 1 + x ) - x + \frac{x^2}{2} \right)
        + \frac{1+x}{1-x} x^2 \ln x\right\},
\end{eqnarray}
with $x \equiv (1-\beta_\mu)/(1+\beta_\mu)$ where $\beta_\mu
\equiv \sqrt{1-4m_\mu^2/s}$. $K(s>4m_\mu^2)$ increases
monotonically from 0.63 to 1 in the range $4m_\pi^2<s<\infty$. The
form of $K$ for $s<4m_\mu^2$ is given in~\cite{AK}, and is used to
evaluate the small $\pi^0\gamma$ contribution to $a_\mu^{\rm
had,LO}$.

To evaluate $\sigma_{\rm had}(s)$, we use experimental data up to
11.09~GeV and perturbative QCD thereafter.
The most important contribution to $\sigma_{\rm had}(s)$ comes
from the $e^+e^- \to \pi^+\pi^-$ channel; the channels
$\pi^+\pi^-\pi^0$, $K^+K^-$, $K_S^0 K_L^0$,
$\pi^+\pi^-\pi^+\pi^-$, $\pi^+\pi^-\pi^0\pi^0$, etc.\ give
subleading contributions. For example, if we evaluate $\sigma_{\rm
had}(s)$ using the sum of the data for the above 6 exclusive
channels up to $\sqrt{s}= 1.43$GeV, we obtain 87\% of the total
$a_\mu^{\rm had,LO}$, with the $\pi^+\pi^-$ contributing 72\%.
Then if from 1.43 to 11.09~GeV we were to use the measurements
$R(s) \equiv \sigma_{\rm had}(s)/\sigma(e^+e^- \to \mu^+\mu^-)$ to
determine $\sigma_{\rm had}(s)$, we would obtain another 11\% of
the total.

As mentioned, our study to update $a_\mu^{\rm had,LO}$ is
motivated by the increasing accuracy of the experimental value for
$a_\mu^{\rm exp}$, see eq.~(\ref{eq:BNL2001}). The special
features of our analysis are (i)~that it is data-driven, based on
all available data, including the new data on exclusive channels
from Novosibirsk, particularly $\pi^+\pi^-$~\cite{CMD2}, and the
BES data on the $R$ ratio~\cite{BES}, (ii)~ the careful
application of a clustering method, so that data of differing
precision can be combined consistently, and (iii)~the use of QCD
sum rules to resolve an apparent discrepancy between the inclusive
and exclusive determination of $\sigma_{\rm had}(s)$ for
$1.4\lesim\sqrt{s}\lesim 2$~GeV.

We chose not to use data on hadronic $\tau$ decays to further
constrain the $e^+e^-\ra2\pi,4\pi$ channels for $\sqrt{s}\lesim
m_\tau$, because of the possible uncertainties connected with
isospin-breaking effects. A careful study of the effects of
including $\tau$ data has been made in Ref.~\cite{DEHZ}.

\section{Processing the hadronic data}
\label{sec:clustering}

We apply the hadronic vacuum polarisation corrections given by
Swartz~\cite{SWARTZ}. Moreover, we calculated the final state
radiative effects for all $e^+e^-\ra\pi^+\pi^-$ data, except for
the new CMD--2 data~\cite{CMD2}, based on eq.~(45) of
Ref.~\cite{HGJ}. These two corrections increase the $\pi^+\pi^-$
contribution by about $1.1\times10^{-10}$, to which we assign a 50\%
error. For the dominant CMD--2 data the radiative corrections are
already included by simply taking the cross section numbers for
$\sigma_{\pi\pi(\gamma)}^0$. Since the recent accurate
data~\cite{OMPI} in the $\omega$ and $\phi$ resonance regions have
not been corrected for any vacuum polarisation effects (see the
comment in~\cite{DEHZ}), we apply the full (including the lepton
vacuum polarisation) corrections to these data. For the data on
the other exclusive channels, and the inclusive data not discussed
in~\cite{SWARTZ}, insufficient information is available to make
reliable radiative corrections. We therefore assign an additional
$\pm1\%$ uncertainty to their contribution to $a_\mu^{\rm
had,LO}$. The net effect is an error of about $\pm2\times10^{-10}$
due to radiative corrections; a further discussion will be given
in~\cite{HMNT}.

We now come to the important problem of `clustering' data from
different experiments (for the same hadronic channel). To combine
all data points for the same channel which fall in suitably chosen
(narrow) energy bins, we determine the mean $R$ values and their
errors for all clusters by minimising the non-linear $\chi^2$
function
\begin{equation}
\chi^2(R_m, f_k) =
 \sum_{k=1}^{N_{\rm exp}} \left[\left(1-f_k\right)/
 {\rm d}f_k\right]^2 + \sum_{m=1}^{N_{\rm clust}}\ \sum_{i=1}^{N_{\{k,m\}}}
 \left[\left(R_i^{\{k,m\}} - f_k  R_m\right)/
{\rm d}R_i^{\{k,m\}}\right]^2. \label{eq:chisquaredef}
\end{equation}
Here $R_m$ and $f_k$ are the fit parameters for the mean $R$ value
of the $m^{\rm th}$ cluster and the overall normalization factor
of the $k^{\rm th}$ experiment, respectively.  $R_i^{\{k, m\}}$
and ${\rm d}R_i^{\{k, m\}}$ are the $R$ values and errors from
experiment $k$ contributing to cluster $m$. For ${\rm d}R_i^{\{k,
m\}}$ the statistical and point-to-point systematic errors are
added in quadrature, whereas ${\rm d}f_k$ is the overall
systematic error of the $k^{\rm th}$ experiment. Minimization of
(\ref{eq:chisquaredef}) with respect to the $(N_{\rm exp} + N_{\rm
clust})$ parameters, $f_k$ and $R_m$, gives our best estimates for
these parameters together with their error correlations.

Our definition (\ref{eq:chisquaredef}) implies piecewise constant
$R$ values but imposes no further constraints on the form of the
hadronic cross section (`minimum bias').  Due to use of the
overall normalization factors it also results in an adjustment of
the different sets within their systematic uncertainties.  This
means, for example, that sparse but precise data will dominate the
normalization. Still, the information on the shape of $R$ from
sets with larger systematic uncertainties is preserved, and all
data contribute weighted according to their significance.

The error estimate for each hadronic channel is then done using
the complete covariance matrix returned by our $\chi^2$
minimization. Therefore statistical and systematic (point-to-point
as well as overall) errors from the different sets are taken into
account including correlations between different energies
(clusters). The minimum $\chi^2$ directly reflects the quality of
the fit and the consistency of the data.  We have checked that for
all hadronic channels we find a stable value and error for
$a_{\mu}^{\rm had,LO}$, together with a good\footnote{In the channels
$e^+e^-\ra\pi^+\pi^-\pi^+\pi^-,\ \pi^+\pi^-\pi^0$, in which data sets
are mutually incompatible, $\chi^2$/degree of freedom
= 2.1,\ 1.6. For both cases the error is enlarged by a factor
of $\sqrt{\chi^2/{\rm dof}}$.}
$\chi^2$ if we vary the minimal cluster size around our chosen
default values (which are typically about 0.2~MeV for a narrow
resonance and about 30~MeV for the continuum).

\begin{figure}
\begin{center}
{\epsfxsize=13cm \leavevmode \epsffile[80 270 490 550]{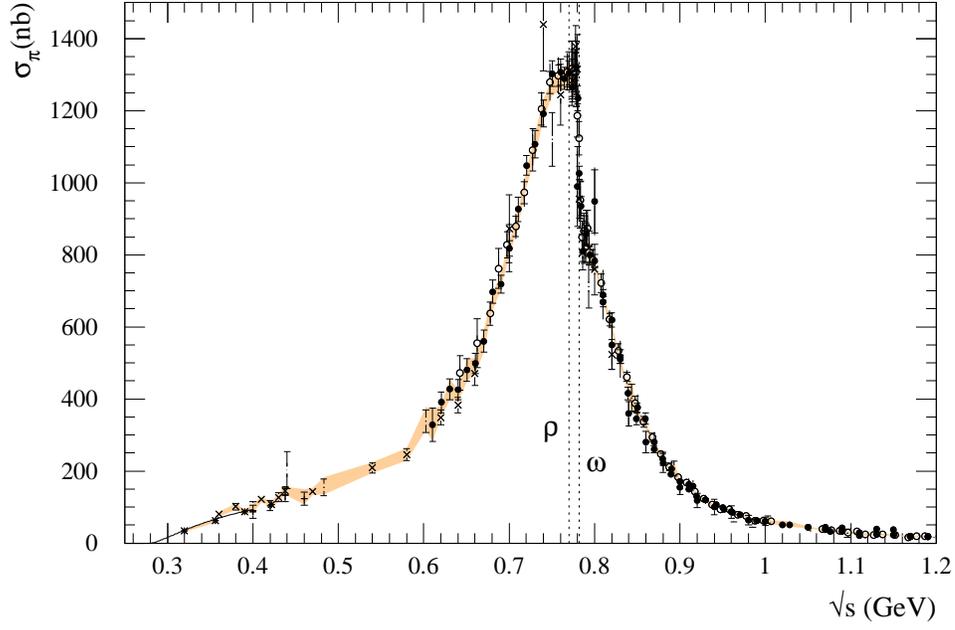}}
{\epsfxsize=13cm \leavevmode \epsffile[65 230 490 555]{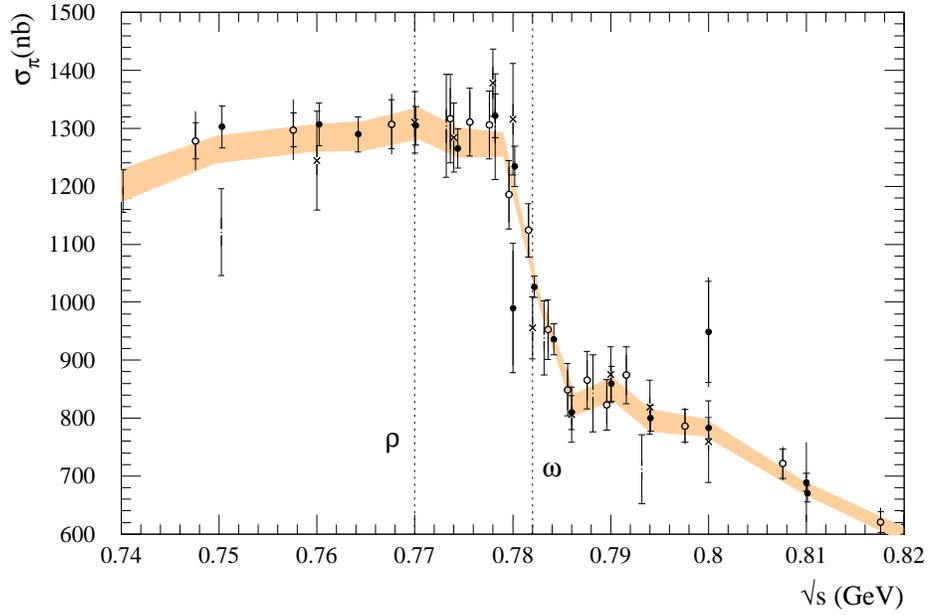}}
\end{center} \vspace{-1cm}
\caption{$e^+e^-\ra\pi^+\pi^-$ data up to 1.2~GeV, where the
shaded band shows the result of clustering. The second plot is an
enlargement of the $\rho$-$\omega$ interference region.}
\label{fig:rhoomega}
\end{figure}
The dispersion integral (\ref{eq:dispersion_rel}) is performed
integrating (using the trapezoidal rule) over the clustered data
directly for all hadronic channels, including the $\omega$ and
$\phi$ resonances.  Thus we avoid possible problems due to missing
or double-counting of non-resonant backgrounds, and interference
effects are taken into account automatically. As an example we
display in Fig.~\ref{fig:rhoomega} the most important $\pi^+
\pi^-$ channel, together with an enlargement of the region of
$\rho$--$\omega$ interference.

\begin{figure}
\begin{center}
{\psfig{file=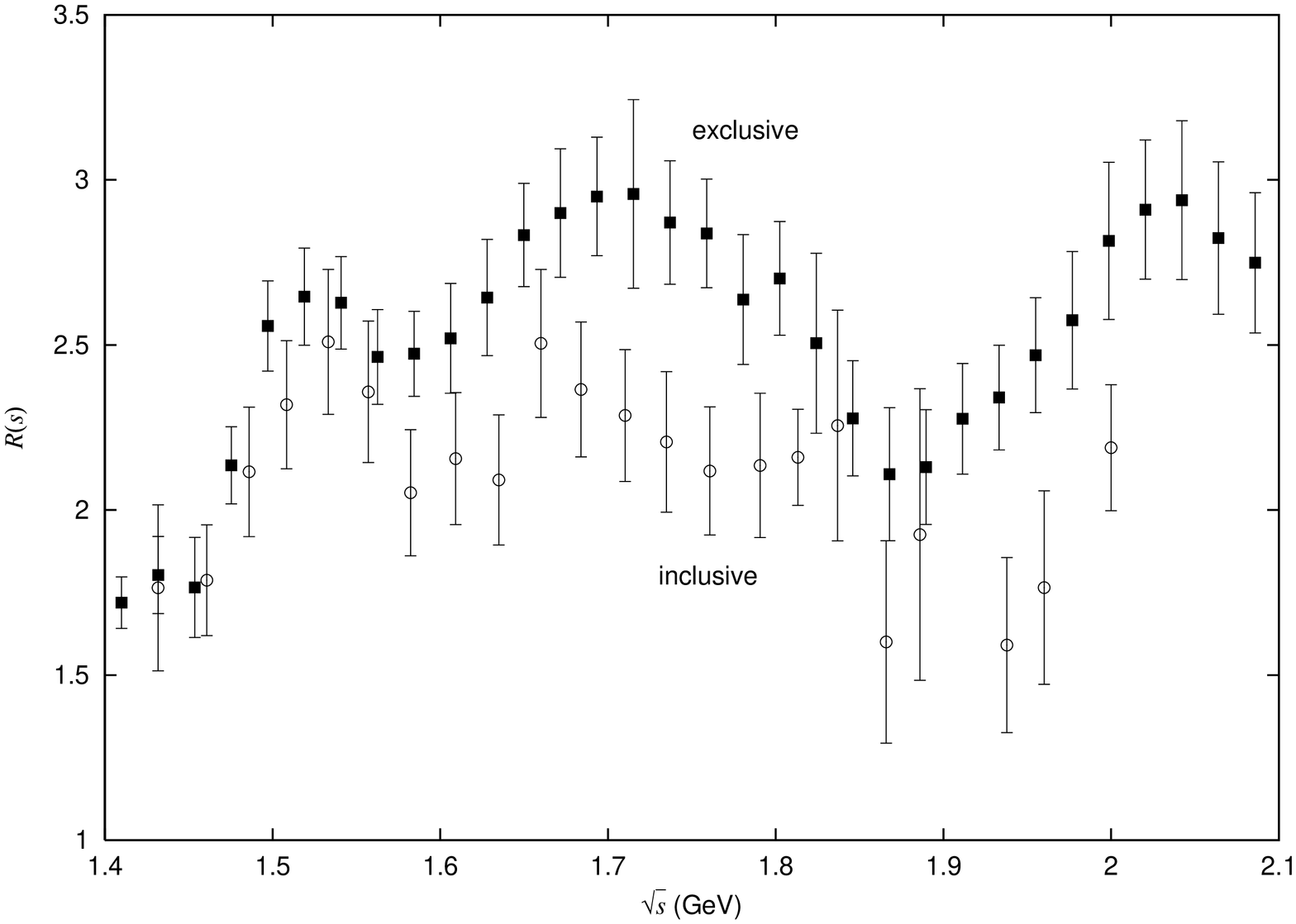,width=8cm,
 bbllx=50pt,bblly=130pt,bburx=550pt,bbury=680pt,angle=-90}}
\end{center}
\caption{The inclusive and the sum of exclusive channel values of
$R$, {\em after} the data have been clustered.}
\label{fig:incl-excl}
\end{figure}

In the region between $1.43$ and $\sim 2$ GeV we have the choice
between summing up the exclusive channels or relying on the
inclusive measurements from the $\gamma\gamma 2$, MEA, M3N and
ADONE experiments \cite{incllow}.  Surprisingly, the sum of the
exclusive channels overshoots the inclusive data, even after
having corrected the latter for missing two-body and (some) purely
neutral modes.  The discrepancy is shown in
Fig.~\ref{fig:incl-excl}, where we display data points with errors
after application of our clustering algorithm.

\section{Results}
\label{sec:results}
\begin{table}[htb]
\begin{center}
\begin{tabular}{|c|c|c|}
\hline
energy range (GeV) & comments & $a^{\rm had, LO}_\mu \times 10^{10}$ \rule[-1ex]{0ex}{3.5ex} \\
\hline
 $ 2m_{\pi} \ldots 0.32 $ & chiral PT & $ 2.30 \pm 0.05 $ \\
 $ 0.32 \ldots 1.43 $ & excl.\ only & $ 596.73 \pm 5.18 $ \\
 $ 1.43 \ldots 2.00 $ & excl.\ only & $ 38.14 \pm 1.72 $ \\
                      & incl.\ only & $ 32.43 \pm 2.46 $ \\
 $ 2.00 \ldots 11.09 $ & incl.\ only & $ 42.09 \pm 1.25 $ \\
 $ J/\psi$ and $\psi(2S)$ & nar. width & $ 7.31 \pm 0.43 $ \\
 $ \Upsilon(1S-6S) $ & nar. width & $ 0.10 \pm 0.00 $ \\
 $  11.09 \ldots \infty $ & pQCD & $ 2.14 \pm 0.01 $ \\
\hline
 $ \sum $ of all & ex-ex-in & $  688.81 \pm 6.17 $ \\
                 & ex-in-in & $  683.11 \pm 5.89 $  \\
\hline
\end{tabular}
\label{tab:hadroniccontr} \caption{A breakdown of the
contributions to different intervals of the integration
(\ref{eq:dispersion_rel}) for $a_\mu^{\rm had,LO}$. The
alternative numbers for the interval $1.43<\sqrt{s}<2$~GeV
correspond to using data for either the sum of the exclusive
channels or the inclusive measurements, see Fig.~2. The total also
includes a small $0.13\times10^{-10}$ contribution from the
$\pi^0\gamma$ channel near its threshold (also included in the second 
line above).}
\end{center}
\end{table}
In Table~1 we list the contributions to $a_{\mu}^{\rm had,LO}$
from different energy regimes: From the two-pion threshold up to
0.32 GeV, chiral perturbation theory is applied (see e.g.\
\cite{CPTreview}--\cite{CPT2}) and for the high energy tail above 11.09 GeV, 
$R$ is calculated using perturbative QCD. The $J/\psi$, $\psi(2S)$ and
$\Upsilon(1S-6S)$ resonance contributions are evaluated in
narrow-width-approximation.  Apart from those contributions we use
the direct integration of clustered data as described above. For
the controversial region from 1.43 to 2 GeV we present two
results: if we use the lower lying inclusive data the
corresponding contribution is considerably smaller than the one
resulting from the use of the sum over the exclusive channels.

A more detailed breakdown of the contributions, and a full
discussion of the data used, will be presented
elsewhere~\cite{HMNT}. There, we will also present an updated
value of the QED coupling at the $Z$ pole, $\alpha(M_Z^2)$.

\section{Resolution of the ambiguity: QCD sum rules}
\label{sec:QCDsumrules}

To resolve the ambiguity between the exclusive and inclusive data
values for $R(s)$ in the range $1.43<\sqrt{s}<2$~GeV (see
Fig.~\ref{fig:incl-excl}), we evaluate QCD sum rules of the form
\begin{eqnarray}
  \int_{s_{\rm th}}^{s_0}\ {\rm d}s\ R(s)f(s)\ =\ \int_C\ {\rm d}s\ D(s)g(s)
  \label{eq:QCDsumrule}
\end{eqnarray}
where $s_0$ is chosen just below the open charm threshold and $C$
is a circular contour of radius $s_0$. $D(s)$ is the Adler $D$
function,
\begin{eqnarray}
  D(s)\ \equiv\
  -12\pi^2s\frac{{\rm d}}{{\rm d}s}\left(\frac{\Pi(s)}{s}\right)\qquad {\rm
  where}\quad R(s)\ =\ \frac{12\pi}{s}\,{\rm Im}\, \Pi(s).
  \label{eq:AdlerDfunction}
\end{eqnarray}
We use the experimental data\footnote{The $J/\psi$ and $\psi(2S)$
resonance contributions are, of course, omitted.} for $R(s)$ (or
equivalently $\sigma_{\rm had}(s)$) to evaluate the
left-hand-side, while QCD is used to determine~\cite{CKK}
\begin{eqnarray}
  D(s)\ =\ D_0(s) + D_{\rm m}(s) + D_{\rm np}(s),
  \label{eq:}
\end{eqnarray}
where $D_0$ is the $O(\alpha_S^3)$ massless, three-flavour QCD
prediction, $D_{\rm m}$ is the (small) quark mass correction and
$D_{\rm np}$ is the (very small) contribution of the condensates.
We take $f(s)$ to be of the form $(1-s/s_0)^n(s/s_0)^m$, with
$n+m=0,1$ or $2$. Once $f(s)$ is chosen, the functional form of
$g(s)$ is readily evaluated. For example, the $n=1, m=0$ sum rule
is
\begin{eqnarray}
  \int_{s_{\rm th}}^{s_0}\,{\rm d}s\;R(s)\left(1-\frac{s}{s_0}\right)\
  =\
  \frac{i}{2\pi}\int_C\,{\rm d}s\;\left(-\frac{s}{2s_0}+1-\frac{s_0}{2s}\right)D(s).
  \label{eq:n=1m=0sumrule}
\end{eqnarray}
The sum rules with $n=1$ or 2 and $m=0$ are found to maximize the
fractional contribution of the left-hand-side of
(\ref{eq:QCDsumrule}) coming from the relevant
$1.43<\sqrt{s}<2$~GeV interval. The evaluations of these two sum
rules are shown in Table~2. Consistency clearly selects the
inclusive, as opposed to the exclusive, determination of $R(s)$.
\begin{table}[htb]
\begin{center}
\begin{tabular}{| c | c | c |}\hline
sum rule & l.h.s. (data) & r.h.s. (QCD) \\\hline
 \raisebox{-1.5ex}[0ex][1ex]{$n=1, m=0 $} & $15.34\pm0.39 $ (incl) & \raisebox{-1.5ex}[0ex][1ex]{$ 15.34 \pm 0.08 $} \\
 & $15.99 \pm 0.35$ (excl) & \\ \hline
 \raisebox{-1.5ex}[0ex][1ex]{$n=2,m=0$} & $10.40 \pm 0.25$ (incl)  & \raisebox{-1.5ex}[0ex][1ex]{$10.30 \pm 0.06$}\\
 & $10.90 \pm 0.22$ (excl) & \\ \hline
\end{tabular}
\label{tab:sumruleresults} \caption{The results of evaluating sum
rule (\ref{eq:n=1m=0sumrule}) and the corresponding one with
$f(s)=(1-s/s_0)^2$, where $\sqrt{s_0} = 3.7$ GeV. 
The main QCD error comes from
$\alpha_S(M_Z^2)=0.117\pm0.002$~\cite{PDG}. The `incl' and `excl'
alternatives refer to using the inclusive or exclusive $e^+e^-$
data in the region $1.43<\sqrt{s}<2$~GeV, see Fig.~2.}
\end{center}
\end{table}
A more detailed discussion of the QCD sum rules, and their
evaluation, will be given in~\cite{HMNT}.

The same conclusion with regard to the resolution of the
inclusive/exclusive ambiguity in the $1.43<\sqrt{s}<2$~GeV
interval was reached in an independent analysis~\cite{MOR}.

\section{Conclusions}
\label{sec:Conclusions}
We have undertaken a data-driven determination of the hadronic
vacuum polarisation contribution to $a_\mu^{\rm had,LO}$. We have
used all available $e^+e^-$ data and a non-linear $\chi^2$
approach to cluster data for the same channel in narrow bins. In
particular, the fit allows the normalizations of the individual
data sets to be collectively optimized within their uncertainties.
We found that there was a discrepancy between the inclusive value
for $\sigma(e^+e^-\ra{\rm hadrons})$ and the sum of the exclusive
channels in the region $1.4\lesim \sqrt{s}\lesim 2$~GeV, which gave an
uncertainty of about $6\times10^{-10}$ in $a_\mu^{\rm had,LO}$. We
used a QCD sum rule analysis to resolve the discrepancy in favour
of the inclusive data. Thus finally we find that the SM predicts
\begin{equation}
a_\mu^{\rm had,LO}\ =\ (683.1\,\pm5.9_{\rm exp}\pm2.0_{\rm rad}
)\times10^{-10}. \label{eq:finalSMprediction}
\end{equation}

Summing up all SM contributions to $a_{\mu}^{\rm SM}$ as given in
eqs.~(\ref{eq:amusm})--(\ref{eq:hadlbyl}), with
(\ref{eq:hadLO_byDH98b}) replaced by (\ref{eq:finalSMprediction}),
we conclude that
\begin{equation}
a_\mu^{\rm SM}=(11659166.9\pm7.4)\times10^{-10},
\end{equation}
which is $36.1\times10^{-10}$ ($3.3\sigma$) below the world average 
experimental measurement. If, on the other hand, we were to take the value of
$a_\mu^{\rm had,LO}$ obtained using the sum of the exclusive data
in the interval $1.43<\sqrt{s}<2$~GeV then we would find
$a_\mu^{\rm SM}=(11659172.6\pm 7.7)\times10^{-10}$, which is 
$30.4\times10^{-10}$ ($2.7\sigma$) below $a_\mu^{\rm exp}$.

An independent SM prediction has very recently been
made~\cite{DEHZ}. Their final $e^+e^-$-based result,
$(684.7\pm6.0_{\rm exp}\pm3.6_{\rm rad})\times10^{-10}$, is very
similar to ours. However, the overall agreement hides larger
differences in individual contributions (but within the quoted
uncertainties).  Our result (\ref{eq:finalSMprediction}) agrees also fairly 
well with a recent reevaluation of the leading hadronic contribution from 
F.~Jegerlehner, who also used the recent CMD-2 data \cite{CMD2} and obtained 
$(688.9 \pm 5.8)\times 10^{-10}$, see \cite{FJ02}. 
In order to facilitate a comparison with these two 
predictions, we will present a detailed breakdown of our result 
elsewhere~\cite{HMNT}.

For the future, we can expect further improvement in the
accuracy of the experimental $(g-2)/2$ measurement.  As far as
the SM prediction is concerned, we may anticipate low energy
data for a variety of $e^+e^-$ channels, produced via
initial state radiation, at the $\phi$-factory DA$\Phi$NE \cite{Kloe} and at 
the $B$-factories, BaBar and Belle, see, for example,~\cite{SOL}.  
For instance, by detecting
the $\pi^+\pi^-\gamma$ channel, it may be possible to measure 
$e^+e^-\to\pi^+\pi^-$ as low as about $\sqrt{s}=400$ MeV.
Moreover, CMD--2 measurements of $e^+e^-\to\pi^+\pi^-$ 
have already been made in this low energy region~\cite{EID}.
When these latter data are final, we anticipate that they would improve
the error on $a_\mu^{\rm had,LO}$ by about $1\times 10^{-10}$.

\section*{Acknowledgements}
We thank Simon Eidelman for numerous helpful discussions
concerning the data. We also thank M.~Hayakawa and T.~Kinoshita
for stimulating discussions and the UK Particle Physics and
Astronomy Research Council for financial support.


\vfill

\end{document}